# Bounds on Photon Charge from Evaporation of Massive Black Holes

C Sivaram and Kenath Arun

Indian Institute of Astrophysics

**Abstract:** Photon charge has been of interest as a phenomenological testing ground for basic assumptions in fundamental physics. There have been several constraints on the photon charge based on very different considerations. In this paper we put further limits based on the well known properties of charged black holes and their subsequent evaporation by Hawking radiation and the assumption of charge conservation over this long physical process.

A photon charge, however miniscule, has been of interest for long as a phenomenological testing ground for basic assumptions in fundamental physics. There have been several very tight constraints on the photon charge based on very different considerations like that of the CMBR, gamma ray bursts [1, 2] and magnetic field deflections. A stringent bound [3] of $10^{-32} e$, where e is the electron charge, was put recently based on the phase coherence of electromagnetic extragalactic radiation, based on the fact that electromagnetic waves moving along different paths do not acquire phase difference.

The particle data group [4] in 2006 listed only four bounds on the photon charge, though several limits have been put on the photon mass. Other recent bounds equally (if not more) stringent have been put based on the dark energy dominance at the present epoch and also on the early universe primordial nucleosynthesis [5]. Current interest in millicharged particles [6, 7] is also relevant in this context.

Here we put further limits based on the well known properties of charged black holes and their subsequent evaporation by Hawking radiation. Charged black holes [8] have an upper limit on the electric charge (Q) they can have, which is related to their mass M by: [9]

$$Q \leq \sqrt{G} M \qquad \ldots (1)$$

For a solar mass black hole this works out to be:

$$Q \leq 10^{39} e \qquad \ldots (2)$$

As is well known [10, 11], when a black hole evaporates by Hawking radiation, it emits thermal radiation with a Hawking radiation with a temperature given by:

$$T_H = \frac{\hbar c^3}{8\pi G k_B M} \qquad \ldots (3)$$

giving a typical energy $(\varepsilon_H)$ for the radiation quanta emitted as:

$$\varepsilon_H \approx \frac{\hbar c^3}{GM} \qquad \ldots (4)$$

With the Stefan Boltzmann law, $E = \sigma T^4 \times$ surface area of event horizon, and using equations (3) and (4), the total number of radiation quanta emitted during the lifetime of the black hole works out to be: (i.e. the total number of photons emitted)

$$N_\gamma = \frac{GM^2}{\hbar c} \qquad \ldots (5)$$

(Indeed this corresponds to the entropy $k_B \cdot \frac{GM^2}{\hbar c}$ associated with the black hole)

As far as the external observer is concerned, the total charge released during the evaporation must equal what went into the formation of the black hole, i.e., that given by equations (1) and (2) and assuming this is distributed among all the radiation quanta emitted, equations (1) and (5), give for a photon charge $q_\gamma$,

$$q_\gamma < \frac{Q}{N_\gamma} < \frac{\hbar c}{\sqrt{GM}}$$

Or in units of electron charge,

$$\frac{q_\gamma}{e} < \frac{\hbar c}{e\sqrt{GM}} \qquad \ldots (6)$$

This for a solar mass black hole gives:

$$q_\gamma < 10^{-38} e \qquad \ldots (7)$$

For a supermassive black hole (of a billion solar mass), equation (6) gives:

$$q_\gamma < 10^{-47} e \qquad \ldots (8)$$

The subtle assumption that charge conservation is valid however long the physical process takes has gone into the result.

Again one would expect (in analogy with the universality of electron charge) that all radiation quanta must have same charge (however small).

Then equations (7) and (8) would imply perhaps the most stringent constraint so far on the photon charge. Again in equation (6), $\dfrac{\hbar c}{e}$ is the unit of quantum flux, $\phi_q$ so

$$\frac{q_\gamma}{e} < \frac{\phi_q}{\sqrt{GM}}.$$

The total charge is $\dfrac{\sqrt{GM}}{e}$, so this gives the total number of quanta emitted as $N_\gamma = \dfrac{GM^2}{\phi_q e}$, which agrees with equation (5)!

$N_\gamma$ can also be interpreted as the total number of quantised flux lines (or tubes) lost from the black hole (the black hole cannot trap magnetic flux, so it is in this sense a diamagnetic object like a superconductor, wherein the flux is expelled like in the Meissner Effect). For analogy with neutron stars see reference [12].

Thus equations (7) and (8) obtained from the physics of black hole evaporation may represent the most stringent constrains on the photon charge. (It is to be noted that for massive black holes $T_H$ is so small, that no (massive) particle pairs are emitted so dominant energy loss is by massless photons).